\documentclass[12pt]{article}

\usepackage{sbc-template}
\usepackage{hyperref}
\usepackage{graphicx}
\usepackage{url,hyperref}
\usepackage[table]{xcolor}
\usepackage[utf8]{inputenc}
\usepackage{adjustbox}
\usepackage{float}
\usepackage[T1]{fontenc}
\usepackage{scalefnt}
\usepackage{grffile}
\title{Recommended Guidelines for Effective MOOCs \\ based on a Multiple-Case Study\thanks{Cite this paper as Eduardo Guerra, Fabio Kon, and Paulo Lemos. \emph{Recommended Guidelines for Effective MOOCs based on a Multiple-Case Study}. Technical Report RT-MAC-2021-02, Department of Computer Science, University of São Paulo, 2021}}

\author{Eduardo Guerra\inst{1}, Fabio Kon\inst{2}, Paulo Lemos\inst{3} }

\address{Free University of Bolza-Bolzano (UniBZ)\\
  Bolzano/BZ -- Italy
\nextinstitute
  University of São Paulo (USP)\\
  São Paulo/SP -- Brazil
\nextinstitute
  Faculdades de Campinas (FACAMP)\\
  Campinas/SP -- Brazil
}

\begin{document} 

\maketitle

\textbf{Technical Report RT-MAC-2021-02}

\textbf{Department of Computer Science, University of São Paulo, 2021}

\begin{abstract}
Massive Open Online Courseware (MOOCs) appeared in 2008 and grew considerably in the past decade, now reaching millions of students and professionals all over the world. MOOCs do not replace other educational forms. Instead, they complement them offering a powerful educational tool that can reach students that, otherwise, would not have access to that information. Nevertheless, designing and implementing a successful MOOC is not straightforward. Simply recording traditional classes is an approach that does not work, since the conditions in which a MOOC student learn are very different from the  conventional classroom. In particular, drop out rates in MOOCs are, normally, at least an order of magnitude higher than in conventional courses. In this paper, we analyze data from 7 successful MOOCs that have attracted over 150,000 students in the past years. The analysis led to the proposal of a set of guidelines to help instructors in designing more effective MOOCS. These results contribute to the existing body of knowledge in the field, bring new insights, and poses new questions for future research.
\end{abstract}

\section{Introduction}

The creation, development, and improvement of knowledge produced by higher education institutions have increasingly used the combination of new technologies and innovative pedagogical strategies. The emergence and evolution of MOOCs (Massive Open Online Courses), since 2012, in particular, and the expansion of the use of online education technologies and pedagogy, in general, are situated in this context.

The experience and knowledge of teachers and researchers offering courses in the MOOC format can be enriched when seeking to improve the design and understanding of these courses' life cycle. Once the phase in which the number of MOOCs present on public, private or hybrid educational platforms has grown, the information produced systematically by these platforms creates the possibility of processing and analyzing data on a scale that would not be possible verified only through classroom courses and the conventional ``brick and mortar'' classroom. That brings the possibility of valuable insights and evidence about online education models. It is possible to understand, among other factors, what is the behavior of students who participate in this type of online education modality.

What makes the interest of thousands of students emerge and makes them look for online courses like MOOCs? With interest aroused, what kind of behavior and decision patterns can be seen in students who follow a complete journey of taking these courses, without abandoning them along the way? What about those students who do not complete the courses, which leads them to abandon the courses? How do the various pedagogical and technological resources provided by a diverse set of companies and organizations and their respective business models dedicated to online education, influence students' behavior and final decisions not to give up online courses \cite{sinclair2016student},\cite{redmond2018online}, \cite{yang2018online}? There are a growing demand and supply of online education, resulting from the effects of the processes of accelerating digital transformation, in times of COVID-19, practiced by the vast majority of educational organizations, including institutions of higher education and by people looking for educational and professional improvement \cite{ferdig2020teaching}?
 
This work aims to systematize and analyze data from a diverse set of 7 MOOCs, in the areas of software technology and entrepreneurship, offered by a wide-ranging online education platform (Coursera), in partnership with 3 public higher education institutions in Brazil (ITA, USP and UNICAMP).

Intending to provide useful concrete insights for the design of MOOCs, this paper uses data characterizing the performance and behavior of students in online courses to answer research questions about (a) the primary intended audience of MOOCs, (b) the moment when students abandon a course, (c) the relation between video duration and students attention, and, (d) the number of times a student performs an online activity.

The results from the characterization and analysis of the data challenge a few common misconceptions about MOOC design. They intend to offer a set of recommendations for teachers, researchers, and practitioners who want to develop or improve the design and maintenance of their online education resources.

\section{Literature Review}

To contextualize our work, let us first examine the relevant body of work on the history of online education, evaluation of experiences, and good practices and methods for MOOCs.

\subsection{History of Online Education}

Since the early 20th century, distance learning was a common way of broadening the reach of education. It started with postal mail-in correspondence courses and later grew to reach large audiences via radio and TV broadcasts.

With the popularization of the Internet, online support for in-classroom courses started to be used extensively in the first decade of the new century. Tools such as Moodle\footnote{\url{https://moodle.org}}, a renowned online open-source learning platform, whose version 1.0 was released in 2002, became highly-popular with thousands of installations worldwide. Initially, these tools used to host and organize supporting material for in-classroom courses, such as slides and exercises. However, with the development of the Internet and digital video technologies, instructors started to store the full video of the lectures on the digital system, sometimes making it completely available for free to anyone on the Internet, as in the case of MIT's OpenCourseware initiative launched in 2003 \footnote{\url{https://ocw.mit.edu/about/milestones}}.

Based on these early experiences, online educators realized that learners enjoyed watching the teachers' movements more than static slides and that short videos focused on specific sub-topics were more pleasant to watch than long 60-minute lectures. Gradually, more and more material was migrated to the online system, making the in-classroom activities, sometimes optional. 

When (1) all the teaching material is available online, including exercises and graded assignments, (2) the course is hosted in a platform capable of handling thousands of students, and (3) the course is available to any interested student (paying or not), we can say that we have a Massive Open Online Courseware (MOOC). 

The first MOOC offered widely, \textit{Connectivism and Connective Knowledge course (CCK08)}, was initiated in 2008 at the University of Manitoba with 25 local students and 2200 online participants worldwide \footnote{\url{https://sites.google.com/site/themoocguide/home}}. New courses were based on the philosophy of connectivism and networking. Modern MOOCs offered by major world universities now follow a more behaviorist approach \cite{Daniel2012}.

\subsection{Evaluations of Experiences}

%%Artigos com avaliações de experiências

Jung and Lee conducted a survey with participants from 5 MOOCs in the Korean open learning support system, based on the edX platform and in the fields of psychology, design, humanities, musicology, and physics \cite{jung2018learning}. By analyzing data from 306 respondents, the authors proposed a model for MOOC engagement and persistence. They concluded that the significant factors that affect MOOC engagement are academic self-efficacy (i.e., the learners' self-confidence in their MOOC performance), teaching presence (i.e., the perception that teachers are present either when elaborating the material or online providing feedback), and perception of usefulness. They also concluded that MOOC persistence was mostly influenced by teaching presence and platform ease of use.

A comparative analysis between courses in the MOOC and SPOC (Small Private Online Course) formats of Shijiazhuang Tiedao University (China) is analyzed by Guo \cite{guo2017mooc}. SPOCs appeared in 2013 at the University of California Berkeley and can be considered a variation of MOOCs by applying online teaching resources to "physical" classrooms on a university campus.

The 4 main advantages of MOOCs can be defined as 1. Share quality resources anytime, anywhere; 2. Active learning; 3. Multiple evaluation mechanisms; 4. Instant feedback. The disadvantages of MOOCs are concentrated in the following three factors: (a) The rate of returning classes remains high; (b) The number of students causes heavy burden of the curriculum to teachers; (c) Authority and credibility questioned. The SPOCs advantage factors are: (a) Learning resources are personalized; (b) Emphasis on positive effects of short videos; (c) Prevent hidden guest situations and provide real-time course management. The main disadvantage factor of SPOCs is symmetrical to one of the advantages of MOOCs: a restricted number of students.

According to Guo \cite{guo2017mooc}, MOOC and SPOC are not a substitute for each other, but parallel experiences. A MOOC is best suited for disseminating principles such as inclusion and equity outside the university campus, serving people with a large capacity for self-learning, and adapting to continuing education. A SPOC is a way of education that adapts to the general conditions of learning on campus, applies to the construction of professional skills, and is more adherent to students with low capacity to control their learning trajectories. In the context of higher education learning on a university campus, SPOCs tend to be superior to MOOCs because they present higher rates of student participation and courses completion.

The experience in Taiwan presents some characteristics that can be considered as reference factors for evaluating national online education initiatives worldwide (\cite {yang2017moocs}). The main strategies to create successful MOOCs were based on the design of dynamic and high-quality courses, carefully activities of post-production, and attractive courses for the students.

Dynamic course design includes goals, weekly progress assessments, modular learning units, and partial and final assessments.  To achieve the goals, instructors plan weekly assessments. The weekly modules are composed of learning units produced with high-quality teaching through lectures, simulations, animations, and movie clips.   The post-production features images, subtitles, sound, film editing, and special effects.   Content is designed and executed to create attractive courses with a "sense of realism," where students can feel analog sensations of a conventional "physical" classroom.  

Since 2013, the Ministry of Education in Taiwan has fostered the development of about 500 MOOCs in Chinese with the involvement of 100 educational institutions and the active participation of Taiwanese companies that have sought to improve primary education and the development of specific skills for specific industrial sectors. These results were derived from the interaction of the developed local education and digital learning industry, the provision of content in high-quality online courses, and partnerships between government, multilevel education institutions (including universities), and companies that conceived, developed, and promoted MOOCs in a joint and articulated way.

\subsection{MOOCs Good Practices and Methods}

Based on his ten-year experience teaching web-enhanced classes, distance learning classrooms, and fully online courses, Savery devised a set of recommendations for instructors in online teaching environments, which he called BE VOCAL \cite{Savery2010}, an acronym for Visible, Organized, Compassionate, Analytical and Leader-by-example. He suggests that for achieving an effective online learning experience, instructors must make themselves \textbf{V}isible, and when possible, the same should be enabled for students. All aspects of the course must be wholly and carefully \textbf{O}rganized. Instructors and teaching assistants should have \textbf{C}ompassion for online learners, who often have multiple problems related to their personal lives. Management of the online learning environment should use \textbf{A}nalytic tools to optimize the effectiveness of the learning experience. Finally, the instructor must behave as a \textbf{L}eader-by-example, i.e., everything he/she does in the course should model best practices in teaching \cite{Savery2010}.

An important issue about methods is the standards and quality criteria applied to the develop\-ment and sustainability of MOOCs \cite{jansen2017quality}. In this study, the quality of MOOCs is evaluated in the context of the use of international quality assurance and measurement frameworks. With the methodological support of case studies from The Open University and the FutureLearn platform, the study describes a national or international quality evaluation framework and the Openup quality certificate. This quality label is derived from the E-xcellence label, a traditional approach to ensuring the quality of e-learning and blended learning that has roots in the experience of open and distance learning institutions. Courses with open licensing and strategies to obtain the certification are required to address the concept of open education for students, demonstrating that the certificate presents principles linked to the Open Education movement. MOOCs require quality assurance processes contextualized to the changes brought by e-learning and open education in institutional structures where international partnerships for MOOCs are growing.

\section{Online learning platforms, types, and business models}

For the provision and availability of MOOCs, there is an increasing need for articulation between public and private actors. In this context, the need to establish relationships between educational organizations (whether for profit or not) and companies specializing in online education appears to be a relevant factor. One of the essential points for the characterization of this type of company is identifying and analyzing the types of online learning offered and the modus operandi of their respective business models.

\subsection{Types of online learning}

Online learning platforms offer products and services that encompass the various modalities of online learning: asynchronous online courses, synchronous online courses, and hybrid courses, described below.

\begin{itemize}
\item \textbf{Asynchronous online courses:}  in this modality, courses do not occur in real-time and can be considered the mode most often used by MOOCs (by design). There is no need for the instructor to interact in real-time and online with students; courses can be taken on demand and not necessarily in a given sequence. Given these characteristics, this type facilitates students' access potential and requires students with greater learning autonomy.

\item \textbf{Synchronous online courses:} symmetrically to asynchronous courses, this modality requires that the interaction between instructor (or intermediaries) and students be done in real-time and online, simultaneously and synchronously. In general, MOOCs do not integrate this type of modality as a permanent strategy, although they may eventually use it. This modality favors students who need more support and guidance for their learning and have a lower capacity to conduct it autonomously.

\item \textbf{hybrid courses:} these are courses that combine synchronous and asynchronous education strategies in online environments.
\end{itemize}

\subsection{Measuring student performance}

When it comes to evaluating student performance, as a MOOC might have thousands of learners, instructors must find highly-scalable approaches for assessing student performance. The most commonly used approaches for that are quizzes, peer review, and automatic correction.

\paragraph{Quizzes.}

Almost all online learning platforms support automatic correction of online quizzes composed of multiple-choice or true/false questions. This evaluation type is useful for checking whether the students are following the key points addressed in each course module. For example, in a course composed of multiple short-video classes, the instructor can create a straightforward quiz with four to ten multiple-choice questions for each video. That is useful both for the student, who can have some feedback on whether he/she understands the most relevant concepts in each video, and for the instructor, who can have a general view of whether his course modules are being adequately used by the student and the concepts being assimilated by the students.

However, quizzes are very limited in that they do not exercise writing skills and can hardly assess students' reflective and creative capabilities.

\paragraph{Peer Review.}

An approach to solve the limitations of quizzes and still be scalable is to use crowdsourcing for grading written exercises, i.e., letting the students themselves correct and comment on the assignments completed by their peers. This approach also has the advantage that, by requiring students to evaluate their online colleagues' exercises, learners are forced to use their analytical skills to assess material related to the course. Thus, students learn not only by doing the exercises and being graded but also by correcting and grading their peers' exercises.

\paragraph{Automatic Correction.}

In a few specific cases, such as programming courses and machine-verifiable contents, such as Mathematics, it is possible to develop computer systems that correct student assignments automatically. Even when the assignment is complicated, and there are thousands of students, a paralyzed computer system running on the cloud can provide prompt feedback to students about their performance in exercises. That is typically achieved by developing a testing suite for each assignment, and the student grade is based on the number of tests that are executed successfully. For example, if the programming assignment is to write a computer program to calculate the roots of a second-degree equation. The course team must write a testing suite evaluating whether the program written by the student computes the desired value correctly in most of the relevant cases. The testing suite is then executed in a virtual machine on the cloud each time a new student submits a new version of its assignment. Then, the student receives a report explaining in which cases his/her program succeeded and in which cases it failed so that he/she can try to find the bug, fix it, and resubmit it. The final student grade is usually an average of the assignment grades, based on the passing tests.

Automatic correction is a powerful tool that can provide students with interactive feedback in a highly scalable way. Student evaluations show that this is a feature that is very appreciated by learners. Unfortunately, the mechanism is limited to specific fields. With advances in artificial intelligence and natural language processing, we expect that, in the future, it might apply to other fields as well.

Current research has exploited computational techniques such as pattern matching, syntax checking, and program mutation to find student mistake patterns in automatic correction systems to improve the guidance and feedback that the system gives to each learner \cite{{Phothilimthana2017,Fox2018}}. That can lead to the automated generation of hints for students who are having difficulties resolving their exercises, acting as a personalized tutor.

\subsection{Configuration and typology of business models of the online education market arising from MOOCs}

A business model can be understood, in general terms, as the combination and logical articulation of the different components of a business in order to describe or analyze new or established products and services. Academic studies such as those of Teece \cite{teece2010business} and Chesbrough \cite{chesbrough2010business} currently coexist with works like the Osterwalder and Pigneur with global business and academic impact \cite{osterwalder2010business}. The business model's logic and functionality have been instrumented in the well-known and globally applied business description and visualization tool, the Business Model Canvas (BMC). In summary, it is possible to verify a pattern familiar to most research and tools related to the business model, which includes 1) identifying and developing customers (how to deliver products and services), 2) describing the productive infrastructure (how products and services are made) and cash flow management (how to capture and appropriate financial returns and how to generate profits).

In addition to a series of markets and products that have been described and analyzed through the previously mentioned business model approach, the online education market emerged from the MOOCs presents a set of studies interested in understanding how different companies and organizations operate under different business models \cite{clift2016educational},\cite{epelboin2017moocs}, \cite{kalman2014race}, \cite{rothe2018competition},\cite{wendler2017business}.

A fundamental feature of  MOOCs is that the provision of this online education modality has been able to drive and generate essential changes in the online education market at the global level. The emergence and initial diffusion of courses in the modality of MOOCs were made possible by new players who entered the online education and learning markets attracted by the possibilities of creating startups and new organizations, generally born as spin-offs of world-class universities like Stanford and Harvard, among others. Beginning in 2012, the consortia of higher education institutions, along with established companies and startups, emerge in the online education scenario and provide MOOCs as part of their activities and portfolios of products and services. Consortia of higher education institutions may be seen, in general, as non-profit organizations, but do not prevent the presence of profitable organizations in their composition. Companies and startups generally have their online education activities aimed at generating profits, but they coexist with companies and startups that present social impact concerns in their value propositions. edX, which emerged from the combined experience and expertise of MIT and Harvard University, and the FutureLearn consortium, both of which emerged in 2012, are among the examples of consortia that bring together higher education institutions and universities that have joined their capabilities to form new organizations and build new competencies. Companies such as Coursera and Udacity, also launched in 2012, offer MOOCs in their product and service portfolios and can be considered as spin-offs of Stanford University.

The MOOC markets are composed of different business models and organizational formats. The online education and MOOCs offerings are marked by technology platforms that can be understood as platforms of software technologies provided by companies that support the creation, development, and management of online courses in various modalities. Most of the companies that offer these platforms work through an ecosystem-based organizational format, which can be considered a way to organize and manage the online education market by engaging the different parts involved in the production, development, and consumption of products and services, as partners and clients such as universities, teachers, instructors, sponsors, and students. Digital platforms based on ecosystems rely on partners to generate value from digital platforms based on software.

This configuration, based on higher education institutions, consortium organizations, companies, and startups, presents different business models that contextualize and influence MOOCs' creation and development. The business models in this market can be classified into 3 types: Freemium Services, Professional Degrees, and Corporate Training \cite{rothe2018competition}. These models coexist and have been adopted by online education market companies in different periods. The leading companies and consortia competing through these types of business models are Coursera, Udacity, edX, and FutureLearn \cite{rothe2018competition}.

In the Freemium Services business model, pioneered by Udacity in 2014, the platforms seek to increase the number of students with additional Premium Services by offering value propositions associated with university academic credentials and the promises of higher employability for students. In general, the platforms sell verified course completion certificates to students and charge a commission of less than  100 percent per completed MOOC. Platforms like Coursera share 6 to 15 percent of gross revenue with partner universities. This business model is identified with the social objectives of the MOOC platforms, by ensuring that students have free access to course content and benefit themselves by the reputation of the platform's partners. The changes in this type of model have evolved from individual certificates per course to the certification of a set of different courses.
 
The Professional Degrees model offered first by Udacity in 2014 launch courses geared towards a complete career-related qualification and therefore differed from the Freemium model that promises access to broader, higher-quality knowledge. Tools such as frequent assignments, group projects, and mentoring/coaching services are available as a strategy to ensure learning outcomes. This model is applied to Udacity Nanodegrees, Coursera Specializations, or edX XSeries, with certifications based on more extended periods. For example, Nanodegrees lasts from 6 to 12 months, providing skills needed for careers and jobs such as data scientists, application developers, or machine learning engineers. In order to meet these requirements, the Professional Degrees model's innovations are centered on new partners such as companies and other institutions outside the academic world as a way to achieve vocational learning goals. Companies like Google and Facebook are among the new partners of Udacity to create Nanodegree courses. In terms of pricing, platforms generally charge a monthly subscription fee for students.

The third type of business model, Corporate Training, was initially adopted by edX in late 2014 and served the corporate employee training market. The value proposition focuses on access to world-class expertise from professionals from large partner companies such as Google and Facebook, who train and develop the corporate customers of the platforms through the courses. In this model, client companies pay a fixed monthly fee per user. Typically, the price per user has decreased with the size of the client company.

There is an inevitable convergence of business models due to the few barriers that companies have in imitating their competitors' innovative business models. While this convergence is reasonably predictable, there remain questions about the next changes and innovations in the online education market, which future research will try to understand. An example of those questions is related to the survival conditions and evolution of the MOOCs format in how they are currently formatted and developed.

In the context of the three types of business models described above, universities and higher education institutions are partners and complement the platforms. An important implication for higher education institutions in general and universities, in particular, is that the latter has a wide range of software platform providers capable of hosting content and knowledge produced by universities. This considerable number of platforms in search of profitable business models can be an essential currency of exchange for universities, which can benefit from this diversity and choose the most advantageous technical and commercial choices in their decision-making processes related to the partner platforms. On the other hand, universities remain dependent on the risks and uncertainties related to the business model that should prevail in the online education market. These uncertainties contribute to the low economic return for universities since these institutions still have the potential to obtain fairer participation in the economic benefits generated in the online education markets.

\section{Multiple-Case Study}

This section describes an analysis performed based on the data extracted from 7 courses developed by 3 renowned higher education institutions. All of them were hosted on the Coursera platform and are offered in Portuguese. The main goal of this analysis is to identify phenomena in students behavior that can generate useful insights for course instructors. The following are the courses under study (the real name of the courses were altered for this anonymized version):

\begin{itemize}

\item Computer Science with Python I~\footnote{\url{https://www.coursera.org/learn/ciencia-computacao-python-conceitos}} and Computer Science with Python II~\footnote{\url{https://www.coursera.org/learn/ciencia-computacao-python-conceitos-2}} - The intended audience of these courses are students and professionals that need an introduction to computer science concepts and programming. The classes are mixed with theory and practical programming, and the evaluation uses programming exercises whose correctness are automatically verified by the platform.

\item Object-oriented Programming with Java~\footnote{\url{https://www.coursera.org/learn/orientacao-a-objetos-com-java}}; Test-driven Development~\footnote{\url{https://www.coursera.org/learn/tdd-desenvolvimento-de-software-guiado-por-testes}}; Introduction to Agile Software Development~\footnote{\url{https://www.coursera.org/learn/principios-de-desenvolvimento-agil-de-software}}; and Agile Development with Advanced Java~\footnote{\url{https://www.coursera.org/learn/desenvolvimento-agil-com-java-avancado}} - The intended audience of these courses are students and professionals that have knowledge in computer programming and need to learn more advanced development techniques. There are separated classes with theory and practical programming, and the evaluation uses quizzes and programming exercises whose correctness usage of the target techniques are peer reviewed.

\item Entrepreneurship~\footnote{\url{https://www.coursera.org/learn/empreendedorismo}} - The course approach is based on the entrepreneurial competencies concepts and practices. An additional founding principle of the course is the entrepreneurial ecosystem acting as a drive for the management of the competencies. All competencies are treated with the support of invited experienced entrepreneurs. These invited skillful people describe and discuss their entrepreneurial experiences and how they get their own competencies. The final goal of the course is to offer concepts and tools to students, turning them better prepared to see, perceive, and apply entrepreneurial competencies in favor of their projects and ventures.

\end{itemize}

The data used in this study was extracted from the Coursera platform at January 19th, 2018. The analysis used demographic information, data about course progress and the forum messages. The data from each course was individually extracted using the Coursera native data export feature. The data from all courses were inserted in a database and custom queries were executed to extract information related to the research questions.

\subsection{Research Questions}
\label{rqs}

The research questions of this multiple-case study are related to the student behavior and characteristics. It focus on identifying patterns on the way that the students act during the course. As a result, based on the evidence found, this paper presents recommendations for the instructors that use the knowledge about the students behavior to improve their experience during their courses. 

\begin{itemize}

    \item RQ1 - Is the student profile the one expected based on the primary intended audience?

    \item RQ2 - What are the the main points where the students abandon the course?

    \item RQ3 - What is the relation between the video duration and the number of students that  concludes it?
    
    \item RQ4 - How many times in average a student perform an activity more than one time?

\end{itemize}

Since the courses had different approaches, the answers for the research questions were evaluated for each one, verifying if an identified behavior can be generalized or if they are particular from a group of courses with a given characteristic. 

\subsection{Methodology}

Based on the research questions, queries were elaborated and executed in the database extracted from the Coursera platform. The data extracted was used to generate charts that present the information in a way that they can be reasoned about. 

The following are the data that were used as the basis for the queries:

\begin{itemize}

\item The amount of students that started each activity in a given course;

\item The amount of students that concluded each activity in a given course;

\item The sequence of activities performed by each student in a course;

\item How many times a student performed an activity in a course;

\item Basic student profile in a course;

\end{itemize}

The data extracted was combined and plotted in charts that were used to extract useful insights related to the research questions. It is important to highlight that the authors of this paper are the instructors of the analyzed courses and their deep knowledge about the course content was considered when interpreting the data.

\subsection{Data Characterization}

The courses selected to be analyzed in this study are detailed in Table \ref{course-data}. All of them are offered in the Coursera platform in Portuguese. They are from different Universities and focus on different types of content. There are programming courses, that covers basic, intermediate, and advanced skills, but there is also a course about software development methodology and another on entrepreneurship. 

\begin{table}[]
\centering
\caption{Course basic data.}
\label{course-data}
\begin{tabular}{|l|l|l|c|c|}
\hline
\textbf{Course Name}                                                                       & \textbf{Institution} & \textbf{Focus}                                                 & \textbf{Weeks} & \textbf{Time}  \\ \hline
\begin{tabular}[c]{@{}l@{}} Computer  Science with Python I\end{tabular}  & IME-USP                  & \begin{tabular}[c]{@{}l@{}}Basic\\ programming\end{tabular}    & 9              & 36h                        \\ \hline
\begin{tabular}[c]{@{}l@{}} Computer Science with Python II\end{tabular} & IME-USP                  & \begin{tabular}[c]{@{}l@{}}Intermediate\\ programming\end{tabular}    & 7              & 35h                        \\ \hline
\begin{tabular}[c]{@{}l@{}}Object-oriented Programming \\ with Java\end{tabular}           & ITA                  & \begin{tabular}[c]{@{}l@{}}Intermediate\\ programming\end{tabular}    & 6              & 50h                        \\ \hline
Test-driven Development                                                                    & ITA                  & \begin{tabular}[c]{@{}l@{}}Advanced\\ programming\end{tabular} & 4              & 32h                       \\ \hline
\begin{tabular}[c]{@{}l@{}}Introduction to Agile Software \\ Development\end{tabular}      & ITA                  & \begin{tabular}[c]{@{}l@{}}Software\\ methodology\end{tabular} & 4              & 32h                       \\ \hline
\begin{tabular}[c]{@{}l@{}}Agile Development with \\ Advanced Java\end{tabular}            & ITA                  & \begin{tabular}[c]{@{}l@{}}Advanced\\ programming\end{tabular} & 4              & 32h                       \\ \hline
\begin{tabular}[c]{@{}l@{}}Entrepreneurship and \\ Entrepreneurial Skills\end{tabular}        & UNICAMP              & Entrepreneurship                                               & 7              & 12h                   \\ \hline
\end{tabular}
\end{table}

Table \ref{course-students} shows the number of students that have started the course (i.e., that has completed at least one course activity), the ones that have completed the course with a passing grade, and the percentage of completion. The table shows the data extracted directly from the Coursera platform on 22nd March 2018 and on August 1st, 2020 The table also presents the average student rating, and it is important to highlight that all of them also have a good rating. 

\begin{table}[]
\centering
{\small
\caption{Active students, conclusions and evaluation}
\label{course-students}
\begin{tabular}{|l|c|c|c|c|c|c|c|}
\hline
                      & \multicolumn{3}{c|}{Jan/2018} & \multicolumn{3}{c|}{Aug/2020}      & \\ \hline
\textbf{Course Name} & \textbf{Started} & \textbf{Completed} &  \textbf{\%} & \textbf{Started} & \textbf{Completed} & \textbf{\%} & \textbf{\begin{tabular}[c]{@{}l@{}}Rating\\ (max 5)\end{tabular}}\\ \hline
\begin{tabular}[c]{@{}l@{}}Introduction to Computer \\ Science with Python I\end{tabular}  & 27,859 & 1,133 & 4 & 82,620 & 5,252 & 6   & 4.9                    \\ \hline
\begin{tabular}[c]{@{}l@{}}Introduction to Computer \\ Science with Python II\end{tabular} & 1,828  & 303   & 17& 6,708 & 1,642 & 24    &4.9                  \\ \hline
\begin{tabular}[c]{@{}l@{}}Object-oriented Programming \\ with Java\end{tabular}           & 21,503 & 879   & 4& 32,032& 1,039 & 3  & 4.8                     \\ \hline
Test-driven Development                                                                    & 5,192  & 449   & 8& 7,360 & 505   & 7  & 4.7                     \\ \hline
\begin{tabular}[c]{@{}l@{}}Introduction to Agile Software \\ Development\end{tabular}      & 3,185  & 144   & 4& 5,874 & 189   & 3    & 4.5                   \\ \hline
\begin{tabular}[c]{@{}l@{}}Agile Development with \\ Advanced Java\end{tabular}            & 2,988  & 264   & 8& 4,370 & 290   & 7        & 4.7               \\ \hline
\begin{tabular}[c]{@{}l@{}}Entrepreneurship and \\ Entrepreneurial Skills\end{tabular}     & 12,564 & 1,123 & 9 & 16,419 & 1,620                   & 10    & 4.7                       \\ \hline
\end{tabular}
}
\end{table}

\section{Data Analysis}

This section presents the analysis performed on the extracted data focusing on the questions presented in Section \ref{rqs}. Each next subsection focus to answer a given research question supported by the data extracted. The charts and tables presented to support our conclusions were selected from some of the courses to illustrate the facts observed, however it is important to highlight that all data was considered in the analysis. 

\subsection {RQ1 - Is the student profile the one expected based on the primary intended audience?}

The primary intended audience for the courses was undergraduate students. One of the initial goals was to give access to the students from the institution responsible for the course as a support and additional material to courses in person. An additional goal was to give access to students from other institutions as well. 

Analyzing the data from the courses in the Table \ref{demographics} it is possible to verify that most of the students do not fit in this initial intended audience. Most of them are full-time employees, graduated, and over 25 years old. Evaluating this data, it is possible to conclude that the main audience are professionals that want to learn about new techniques, programming languages, technologies, as well as get advice on business and entrepreneurship. Thus, we see that MOOCs like this can attract a large number of older professionals seeking to recycle their knowledge.

However, the undergraduate students are still a significant part of the participants. We can observe that by the number of students with no higher education, not employed, and bellow 24 years old.  

We can also observe from the data that the vast majority of students are male in all courses. Thus, unfortunately, these courses are not helping to close the gender gap in tech education. The Entrepreneurship course is the one with the largest percentage of female students, 32\%.

\begin{table}[]
\centering
\caption{Demographics from the multiple-case study courses}
\label{demographics}
\begin{tabular}{|l|c|c|c|c|c|c|c|}
\hline
\textbf{Demographic}                                          & \textbf{C1} & \textbf{C2} & \textbf{C3} & \textbf{C4} & \textbf{C5} & \textbf{C6} & \textbf{C7} \\ \hline
\multicolumn{8}{|l|}{\cellcolor[HTML]{C0C0C0}{\color[HTML]{000000} Gender}}                                                                                     \\ \hline
Woman                                                         & 12\%        & 12\%        & 13\%        & 10\%        & 16\%        & 11\%        & 32\%            \\ \hline
Men                                                           & 88\%        & 88\%        & 87\%        & 90\%        & 83\%        & 89\%        &  68\%           \\ \hline
\multicolumn{8}{|l|}{\cellcolor[HTML]{C0C0C0}Age}                                                                                                               \\ \hline
\textgreater= 24                                              & 28\%        & 30\%        & 24\%        & 15\%        & 16\%        & 16\%        &  15\%           \\ \hline
25-34                                                         & 39\%        & 37\%        & 43\%        & 50\%        & 40\%        & 47\%        & 40\%            \\ \hline
35-44                                                         & 22\%        & 22\%        & 24\%        & 29\%        & 30\%        & 26\%        & 27\%            \\ \hline
\textgreater45                                                & 11\%        & 11\%        & 9\%         & 6\%         & 14\%        & 11\%        &  18\%           \\ \hline
\multicolumn{8}{|l|}{\cellcolor[HTML]{C0C0C0}Education}                                                                                                         \\ \hline
Doctorate                                                     & 3\%         & 3\%         & 2\%         & 2\%         & 4\%         & 3\%         & 3\%             \\ \hline
Master's                                                      & 16\%        & 16\%        & 12\%        & 16\%        & 20\%        & 15\%        & 21\%             \\ \hline
Graduate                                                      & 43\%        & 43\%        & 47\%        & 57\%        & 51\%        & 56\%        & 50\%             \\ \hline
\begin{tabular}[c]{@{}l@{}}No higher\\ education\end{tabular} & 38\%        & 38\%        & 39\%        & 25\%        & 25\%        & 26\%        & 26\%            \\ \hline
\multicolumn{8}{|l|}{\cellcolor[HTML]{C0C0C0}Employment Status}                                                                                                 \\ \hline
Full-time                                                     & 59\%        & 59\%        & 59\%        & 71\%        & 66\%        & 68\%        & 60\%             \\ \hline
Part-time                                                     & 11\%        & 11\%        & 13\%        & 10\%        & 9\%         & 10\%        &  15\%            \\ \hline
Not employed                                                  & 28\%        & 28\%        & 26\%        & 19\%        & 24\%        & 22\%        & 23\%             \\ \hline
Other                                                         & 2\%         & 2\%         & 2\%         & -           & 1\%         & -           &  2\%            \\ \hline
\end{tabular}
\end{table}

\subsection {RQ2 - What are the the main points where the students abandon the course?}

\begin{figure}[!ht]
	\centering
	\includegraphics[width=\linewidth]{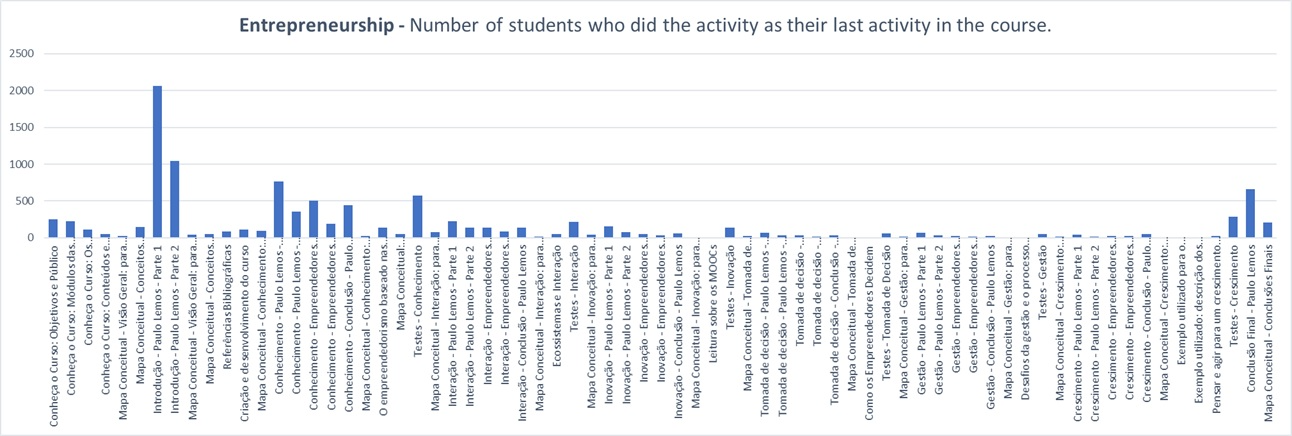}
	\caption{Number of students that dropped out in each activity of the Entrepreneurship course.}
	\centering
	\label{fig:dropout_emp}
\end{figure}

\begin{figure}[!ht]
	\centering
	\includegraphics[width=\linewidth]{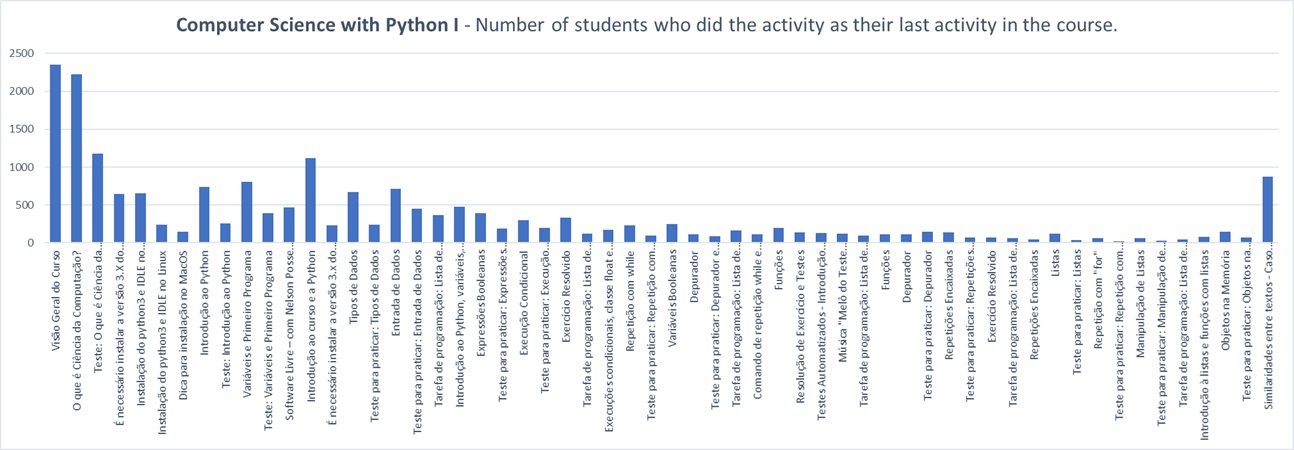}
	\caption{Number of students that dropped out in each activity of the Computer Science with Phython I course.}
	\centering
	\label{fig:dropout_phython}
\end{figure}

\begin{figure}[!ht]
	\centering
	\includegraphics[width=\linewidth]{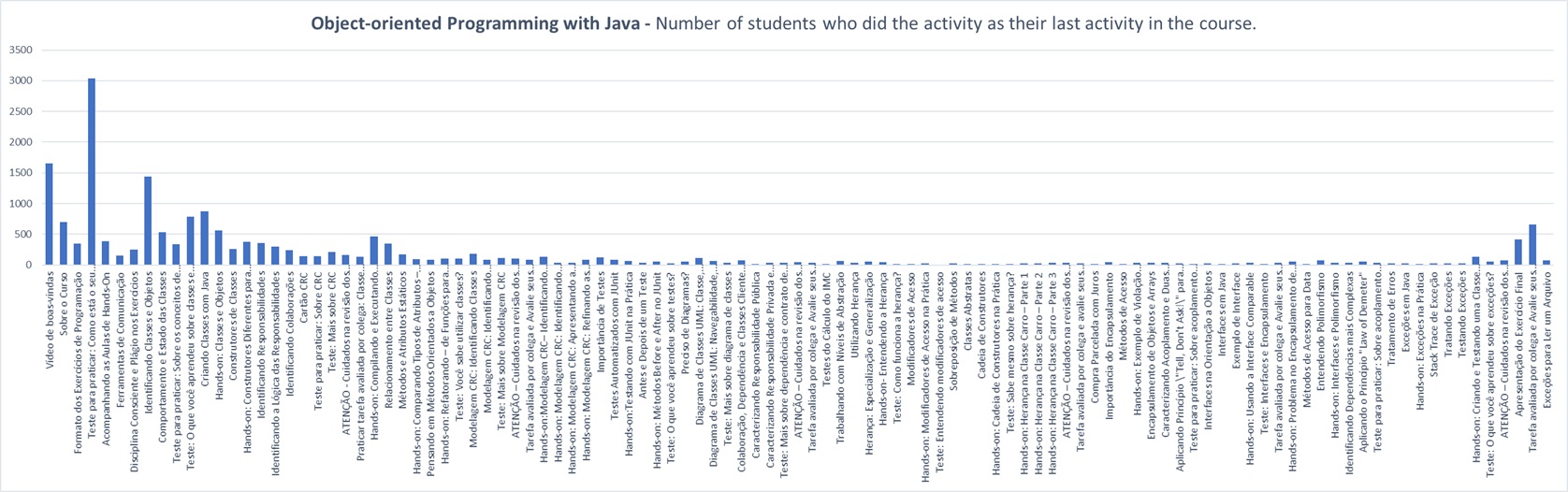}
	\caption{Number of students that dropped out in each activity of the Object Oriented Programming with Java course.}
	\centering
	\label{fig:dropout_java}
\end{figure}

Figures \ref{fig:dropout_emp},  \ref{fig:dropout_phython}, and  \ref{fig:dropout_java} presents charts with the number of students that performed each activity as the last course activity and did not concluded the course. The activities are presented in the same order that they appear in the course. So, it can be considered that the horizontal axis represent the course time line, and the bar represent the number of students that stopped the course in that point. The graphs from the other courses have a similar shape and the conclusions were done based on the analysis of all of them.

The greatest drop out point is in the beginning of the course. This can be considered a natural behavior from potential students that want to get more information about the course and understand its scope. After the initial activities, the number of students that drop out reduces gradually until it reaches a point that the students are engaged in the course and the number is very small.

Another evident drop out point is in the final activities of the course. All the courses analyzed have a final exercise that is harder and more laborious that the others in the course. That can be one of the reasons why this exercise has a high drop out rate. Another explanation can be the fact that some students are interested in absorbing knowledge only from the videos, ignoring the exercises in general along the course. In this scenario, the student performs the course until the end, but does not conclude it.

Another observation that can be done is that even with the drop out rate reducing along the course activities, there are some local peaks. These small peaks happen in the final module activities. This indicates that is more common for a student to drop out the course in the end of a module than in the middle of it. 

\subsection {RQ3 - What is the relation between the video duration and the number of students that concludes it?}

\begin{figure}[!ht]
	\centering
	\includegraphics[width=\linewidth]{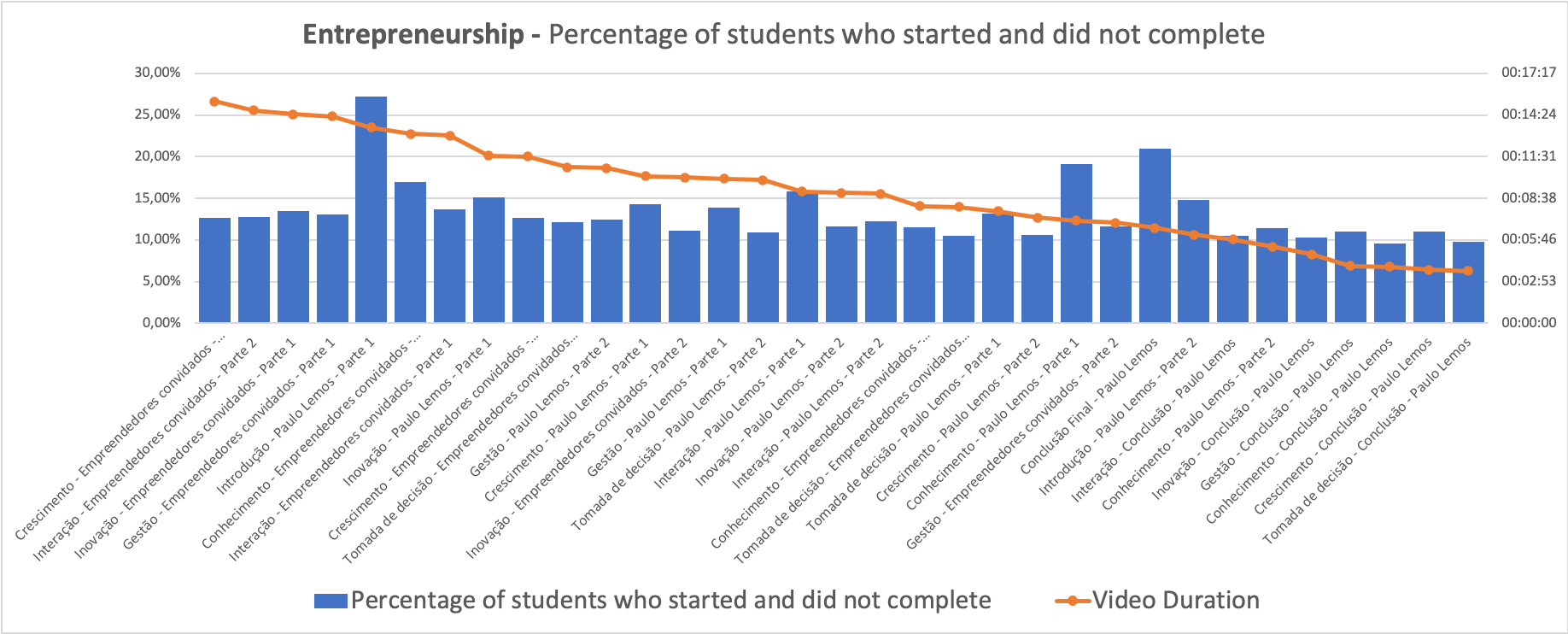}
	\caption{Number of students that started and did not complete each activity of the Entrepreneurship course.}
	\centering
	\label{fig:emp-time_leave}
\end{figure}

\begin{figure}[!ht]
	\centering
	\includegraphics[width=1.0\columnwidth]{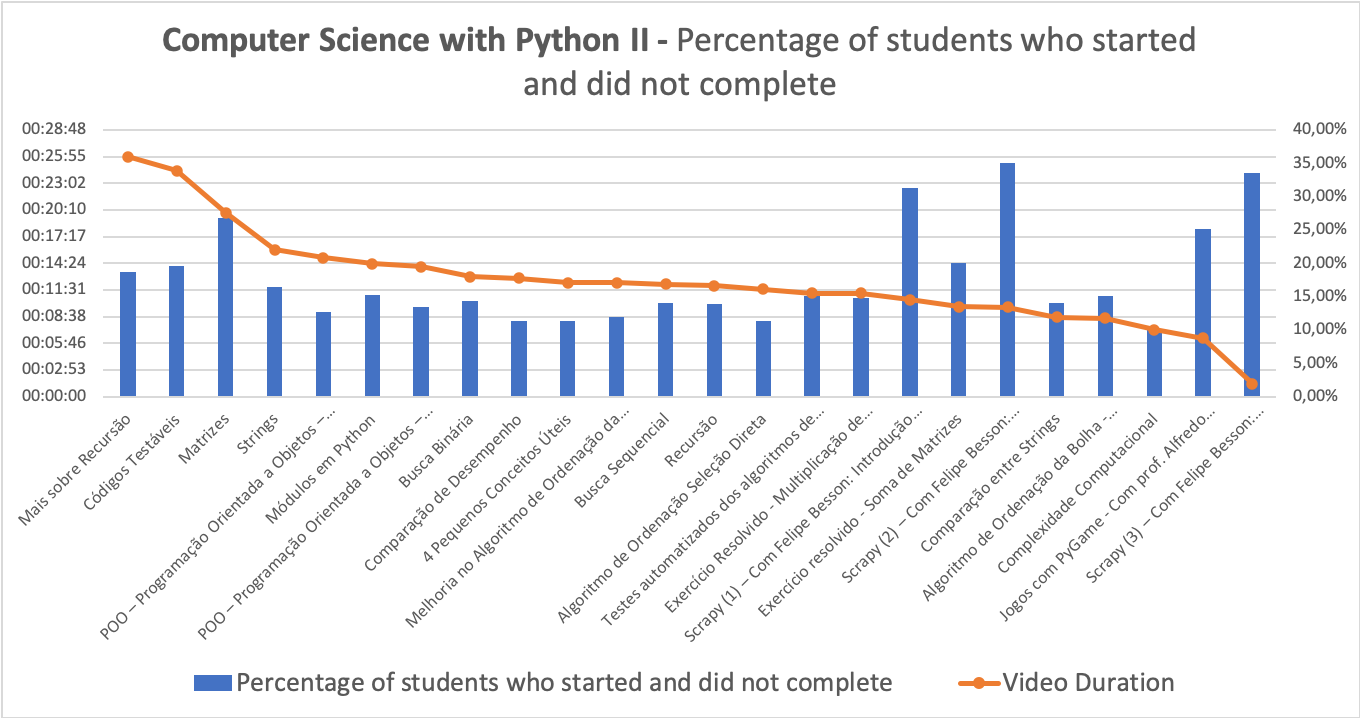}
	\caption{Number of students that started and did not complete each activity of the Computer Science with Phython II course.}
	\centering
	\label{fig:python2-time_leave}
\end{figure}

\begin{figure}[!ht]
	\centering
	\includegraphics[width=1.0\columnwidth]{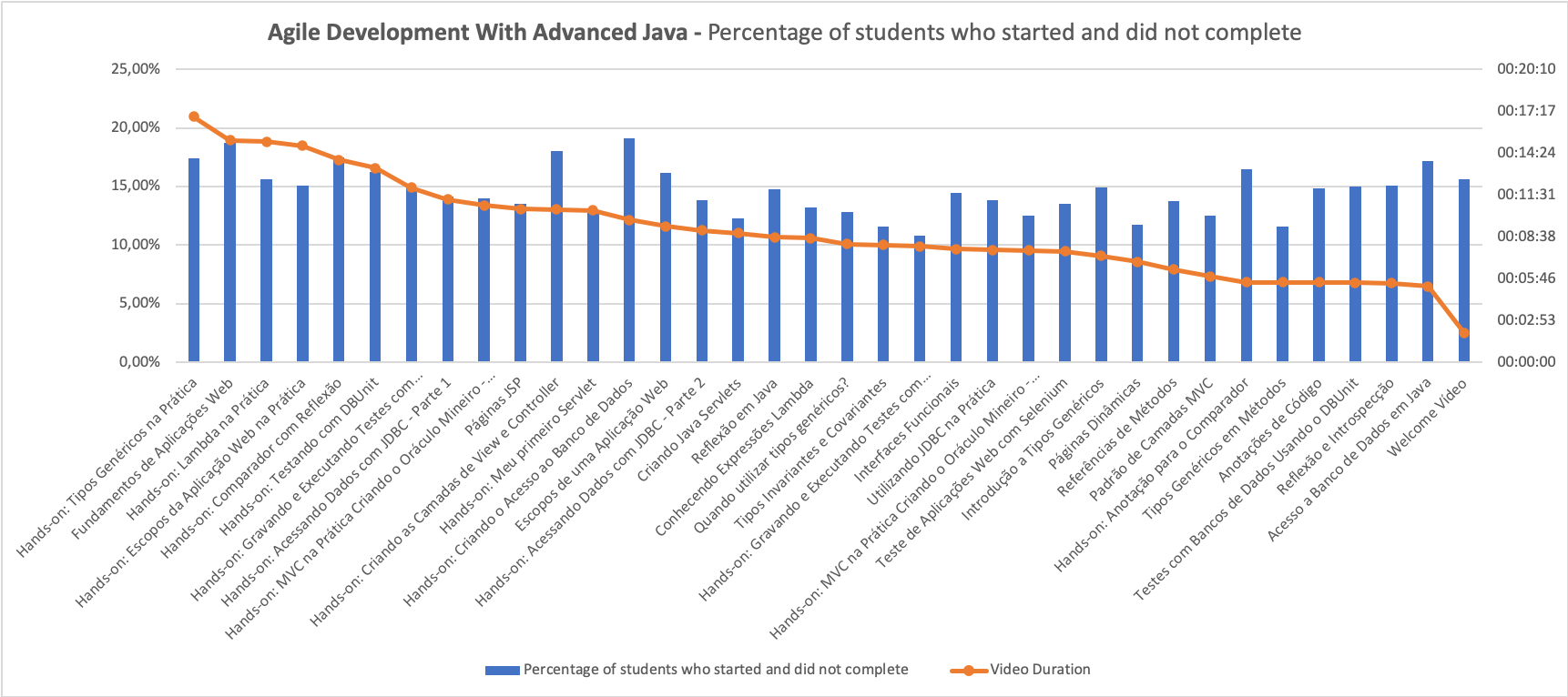}
	\caption{Number of students that started and did not complete each activity of the Agile Development with Advanced Java course.}
	\centering
	\label{fig:advanced-time_leave}
\end{figure}

Figures \ref{fig:emp-time_leave}, \ref{fig:python2-time_leave} and \ref{fig:advanced-time_leave} presents the number of students that dropped out each video activity. In each chart, the activities are ordered based on video total time in decreasing order. For the Entrepreneurship course chart in Figure \ref{fig:emp-time_leave} the videos varied from  03:36 to 15:20; for the Computer Science with Python II in Figure \ref{fig:python2-time_leave} the videos varied from  06:17 to 25:53; for the Agile Development with Advanced Java course chart in Figure \ref{fig:advanced-time_leave} the videos varied from 02:10 to 16:54. 

Based on the charts, it is possible to observe that there is no relationship between the duration of the video and the number of students who dropped out in each of them. This evidence contradicts the widely disseminated idea in MOOC guidelines that larger videos have a higher probability to make the students to drop out the activity. 

For instance, in the Advance Java course, whose chart is presented in Fig. \ref{fig:advanced-time_leave}, most of the longer activities are the ones called as "hands-on" with practical programming exercises. For these kind of videos, to split them in two or more would break the line of thought, making it more difficult for students to follow. Different styles of videos might have a distinct retention from students. Therefore, further research should investigate the influence of the kind of video in the optimal duration time to maximize student retention.

\subsection {RQ4 - How many times, in average, a student performs an activity more than one time?}

\begin{figure}[!ht]
	\centering
	\includegraphics[width=1.0\columnwidth]{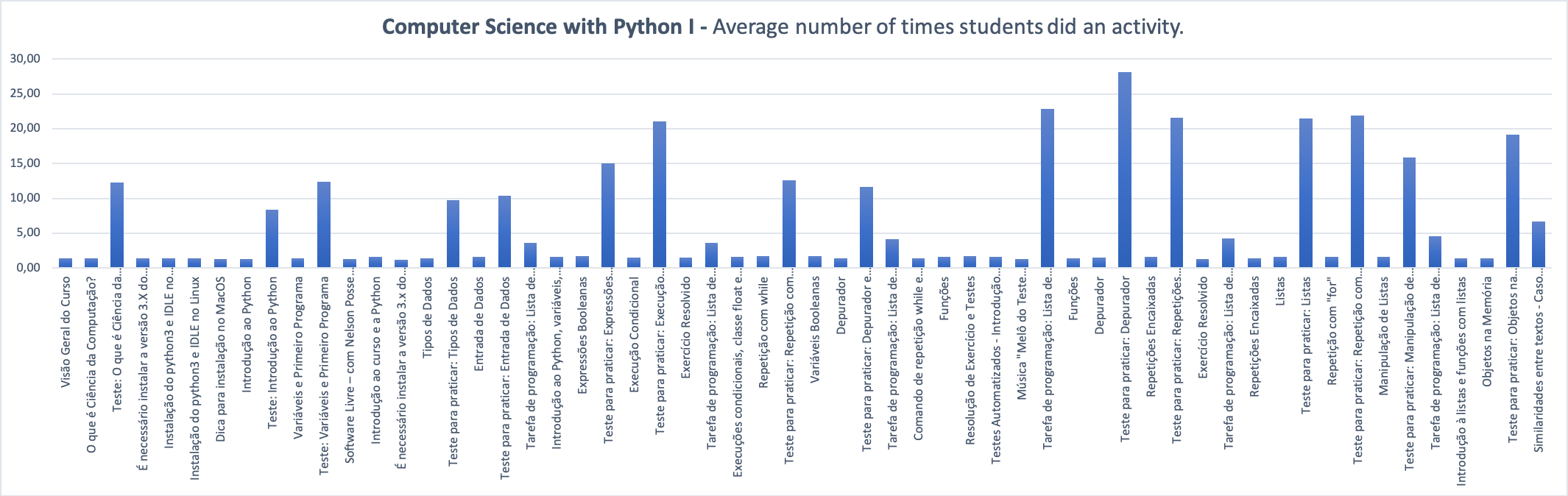}
	\caption{Average Number of Times that Students Performed Each Activity on Computer Science with Python I}
	\centering
	\label{fig:pyth1-avg_num_act}
\end{figure}

\begin{figure}[!ht]
	\centering
	\includegraphics[width=1.0\columnwidth]{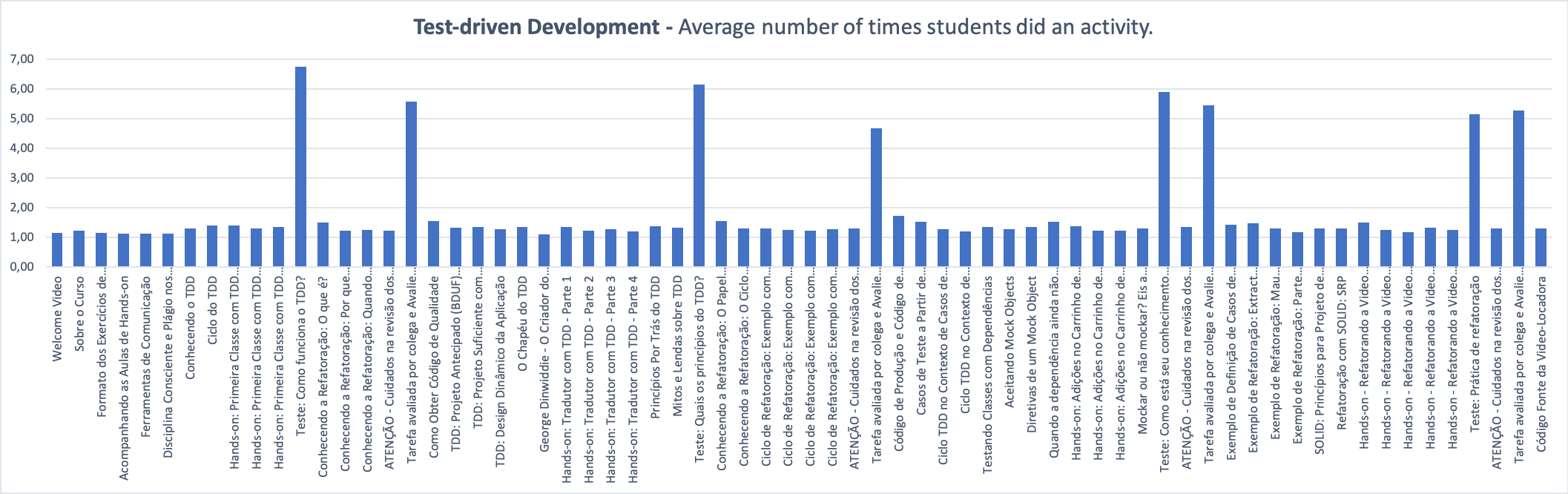}
	\caption{Average Number of Times that Students Performed Each Activity on Test-Driven Development}
	\centering
	\label{fig:tdd-avg_num_act}
\end{figure}

Figures \ref{fig:pyth1-avg_num_act} and \ref{fig:tdd-avg_num_act} present the average number of times that an activity is performed in the Computer Science with Python I and Test-Driven Development courses, respectively. Some activities clearly stands out as being accessed several times by the participants. These are the course evaluation activities, like tests and programming assignments. The same pattern can be found on all the other courses.

\begin{figure}[!ht]
	\centering
	\includegraphics[width=1.0\columnwidth]{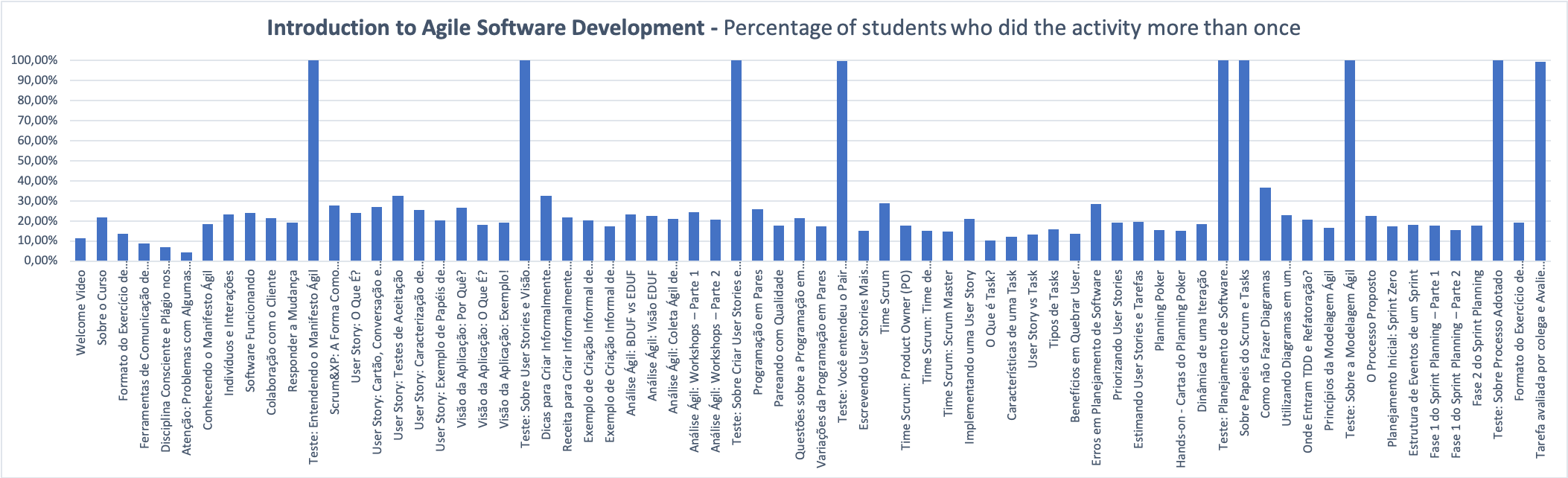}
	\caption{Percentage of Students that Performed an Activity More Than One Time on Introduction to Agile Software Development}
	\centering
	\label{fig:agile-morethanonetime}
\end{figure}

\begin{figure}[!ht]
	\centering
	\includegraphics[width=1.0\columnwidth]{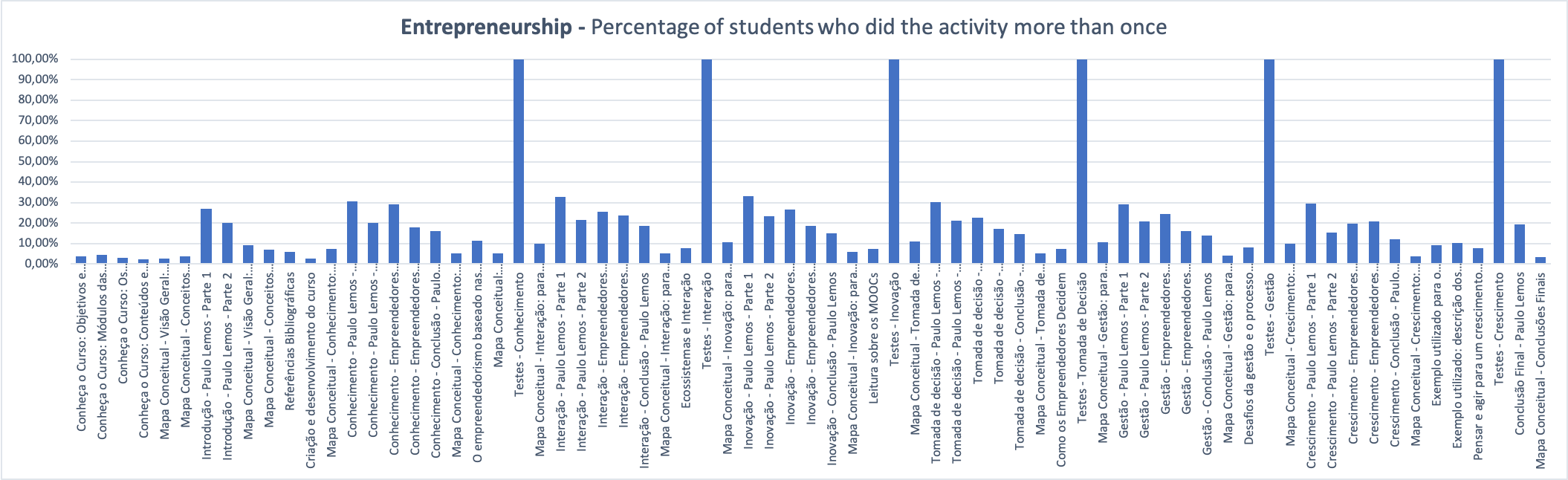}
	\caption{Percentage of Students that Performed an Activity More Than One Time on Entrepreneurship}
	\centering
	\label{fig:ent-morethanonetime}
\end{figure}

Evaluating the data from another point of view, Figures \ref{fig:agile-morethanonetime} and \ref{fig:ent-morethanonetime} show the percentage of students that enter the activity more than one time for two other courses. Excluding the activities aimed for students evaluation, that are accessed more than one time by 100\% of the students, the other ones usually have more than one access from 10\% to 30\% of the students.

Based on that, the data shows that the majority of the students watch the videos only one time, varying from 10\% to 30\% the percentage of the students that repeat it. So, the premise that having the recorded classes available will make the students to access this resource several times is not true for most of them. The instructors should consider that, for the vast majority of students, each video will have only a single opportunity to pass its message.  

\section{Recommendations}

The data analysis for the 7 courses shows some interesting patterns that lead to useful recommendations for instructors to prepare effective MOOCs. In the following, we analyze the major phenomenons that we identified in the data collected from the courses. For each one, we include a discussion and a recommendation that can be followed by instructors to deal with it.

\subsection{High Initial Evasion}

\textbf{Phenomenon description:} MOOCS usually have a very high evasion rate that mostly happens in the first week.

\textbf{Discussion:} Since the entrance barrier to starting a MOOC is low, it is expected that they will have a drop out rate higher than conventional courses. When the course is free, many students enroll in it aiming to look at the introductory videos to have a quick idea if he/she wants to continue. As a result, it is prevalent for MOOCS to have more than 90\% of dropouts and, sometimes, 70\% of the dropout rate in the first week.

\textbf{Recommendations:} Since most of the course audience will only watch the first videos, the instructor should try to captivate them in the first week. In particular, during this period, avoid difficult tasks and concentrate on motivating the student for the course. Show how this course will benefit the student life or career, what he/she will learn, and why this is important. If possible, the instructor should include an important learning objective that the student can already get in the first week.

\subsection{Inter-Module Drop-out}

\textbf{Phenomenon description:} The number of students attending the course diminishes near linearly in each new week (module) of the course.

\textbf{Discussion:} In all the courses analyzed, the number of students decreases every week of the course. Even though these courses were very well evaluated by the students (grades 4.8 or 4.9 out of 5.0), still, the vast majority of the students abandon the course before completing it.  Breaking longer courses into smaller courses make it possible for a more significant portion of the students to complete at least one of the courses and, if they want, to get a certificate. For example, if we break an 8-week course into two 4-week courses, students that are interested in getting just an introduction to the topic can complete the first part of the course (initial 4 weeks) and students that already know the basics and want to jump to the most advanced part have the choice to skip the first course and complete only the second part of the course.

\textbf{Recommendations:} Courses should not be very long, e.g., over 6 weeks. Longer courses should be broken into smaller units, each with 3 to 5 weeks in length. This practice gives more flexibility to the students, promoting a better engagement, and decreasing the total drop-out rate.

\subsection{Intra-Module Drop-out}  

\textbf{Phenomenon description:} Inside each module (typically lasting for 1 week, according to Coursera's recommendations), the number of students attending the course diminishes (following a negative exponential distribution) with each activity in the week. At the beginning of each module, we see a peak in student engagement. The beginning of a new module always brings more engagement than the end of the previous module. In Figures \ref{fig:dropout_emp},  \ref{fig:dropout_phython}, and  \ref{fig:dropout_java}, the activities that present a higher local drop-out rate are in the beginning of the module, which means that more students abandon the course starting a module than in the middle of it. 

\textbf{Discussion:} This phenomenon occurs probably because of a combination of factors. When starting a new module, the student tends to be less tired than when working in a sequence of tasks closer to the end of a module. Also, the excitement of getting in touch with a new topic might bring an energy boost to the student. However, perhaps the most significant factor is that experienced students might not like to follow the course sequentially in the order that the instructor determined, but rather prefer to jump to the topics that he/she believes would be more valuable. Thus, it is common for this kind of student to start the first activities of a module to check whether he/she would have something to learn from it and, in many cases, abandon it and jump directly to the start of another module.

\textbf{Recommendations:} If a module is essential to students, the MOOC designer should add fun and joyful activities in the middle of it to avoid having a drop out at this point. Additionally, it can be stated explicitly to the student why this module is so important. On the other hand, if the instructor wants to benefit from this phenomenon, allowing students to select the topics that he/she is interested in and pay attention only to those topics, the following procedure is recommended: (a) present an overview of each module in the beginning of the course to allow the student to choose the topics of his/her interest; (b) in the beginning of each module present a summary about its content.

\subsection{Retention vs. Video Duration}

\textbf{Phenomenon description:} Previous research indicates that shorter videos provide better retention from online students. A study with MIT MOOCs suggests that 6 minutes is the ideal length of videos \cite{guo2014video}, while Coursera recommends keeping videos shorter than 8 minutes. However, based on the data analyzed, there was no relation between the length of the video and the number of students that dropped it in the middle. 

\textbf{Discussion:} In our courses, we had to include a few short videos to explain how to do some technical activity, which is usually something boring or uninteresting; these kinds of videos had low retention even though their length was short. On the other hand, we had a few long videos whose contents were very engaging, for instance, videos presenting hands-on programming activities.  Overall, as we observe in Figs. \ref{fig:emp-time_leave}, \ref{fig:python2-time_leave}, and \ref{fig:advanced-time_leave}, our data shows absolutely no correlation between the length of the video and the interest demonstrated by students.

\textbf{Recommendations:} When deciding the length of the videos, instructors should not strictly follow a magic number derived from an average of thousands of courses. It is essential to consider the specific nature of the course, the area that it belongs to, and the attention span of the target audience. Also, if the video covers an activity that naturally takes more time or a more complex subject, it is acceptable to go longer.  Breaking a video in two to conform to a given time threshold might have a negative effect on reducing the engagement by cutting the line of reasoning.

\subsection{Video Replays}

\textbf{Phenomenon description:} Only a small number of students, between 10\% to 30\%, watch a video more than once.  

\textbf{Discussion:} In an online course, the fact that students have all the material available and can replay videos, might lead to the conclusion that the lectures should be objective and content should not be redundant. However, our data shows that only a few students watch a video more than once during the course. So, the instructor should not rely on the fact that all the students will watch a video again if a given concept is not clear.

\textbf{Recommendations:} If a concept is essential to be understood by the students, it should be reinforced through different lectures and materials. Few students will repeat a video if something is not clear. So, even if the student can replay the videos, it is good to plan a level of redundancy of the fundamental concepts at different points of each video and of the course.

\subsection{Repetition of Evaluation Tasks}

\textbf{Phenomenon description:} Almost all the students repeat evaluation activities, which also have a high average number of repetitions.

\textbf{Discussion:} Our data shows that, differently from the lecture videos, the students enter several times in the activities that evaluate their performance in the course. This repetition can be due to several attempts to complete the activity. It can also be related to the student behavior to enter the activity to check what knowledge is necessary to complete it.

\textbf{Recommendations:} That information can be used by instructors to reinforce essential concepts in the course. Evaluation tasks are a checkpoint accessed several times by students that want to have a good performance in the course. The review of the concepts that the instructor wants to highlight can be added to the evaluation activity as an introductory text or video, or even in the activity itself.

\subsection{Broad Audience}

\textbf{Phenomenon description:} Courses designed and planned to be an introduction to a particular topic are used not only by beginners but also by experienced professionals to update their knowledge and serve as a bonus in their curriculum while seeking new jobs. 

\textbf{Discussion:} Although the seven courses we analyze in this paper were initially designed to be the first contact of the students with the field; our study showed that around half of the students are full-time employees whoc are seeking to improve their curriculum. Less than 20\% of the participants are full-time students.

\textbf{Recommendations:} When designing an introductory course, be aware that it can be useful not only to people who are seeing that content for the first time but also for professionals who wish to learn more recent approaches and techniques in the field that they already work. Providing continuous education to experienced professionals is especially relevant in fields that advance fast such as Information and Communication Technologies. Thus, it might be a good idea to provide a broad overview of the course contents in the first week so that students can identify the topics in which they have an interest or not and focus their attention on what is more critical for them. Nevertheless, the course designer must be aware that students do not always know the most relevant material to achieve the career goals they have in mind. Thus, educating the students about this is also something that the instructor should do at the very beginning of the course. The instructor should also provide alternative paths that students could take within one course or a collection of courses based on their personal goals.

%Boas práticas (padrões?) para construção de MOOCs

%Tipos de material didático

%Formato de aula (tempo, tamanho, continuidade em várias aulas)
%Tanto teóricas quanto práticas

%Boas práticas de aulas práticas são as mesmas de aulas teóricas?

%Como os alunos acompanham as aulas práticas? Só assistem? Acompanham?

%Tipos de avaliação do desempenho do aluno

%Exercícios de programação com revisão em pares (foco no design de software) - 

%experiência, como é o aprendizado na correção, como é o feedback recebido

\section{Conclusions and Future Directions}

This paper presented a quantitative study based on MOOC data with the objective of deriving guidelines to help MOOC instructors designing better courses. A multiple case study evaluated 7 courses from 3 different Brazilian institutions, extracting data about the behavior of over 150,000 students. 
We analyzed information about the student profile, course drop-out, incomplete activities, and video replays. Based on that, we derived a set of recommendations for instructors to guide MOOC creation inspired on student behavior. 

The business models of companies that offer online education can be crucial in defining the type of audience for MOOCs. The evidence that undergraduate students are not the majority of MOOC students may be related to the fact that the marketing of online education companies is aimed at promoting aspects related to professional training and employability, factors that are not always among the main concerns of undergraduate students, especially those at the beginning of their undergraduate courses.

The article shows that the causes of the attractiveness of content (in this case, video content) may be related more to aspects of the educational quality of the learning material and the teacher than to quantitative aspects (size of videos) or technological platforms.

Our five-year experience with MOOCs show that, in addition to following guidelines such as the ones presented in this paper, it is important to dedicate a substantial amount of time to refine the course after its initial deployment. Thus, we recommend that instructors should monitor student activity so that they can identify what are the strong and weak points of the course. One can identify the strong points by reading the student's stories, comments, and likes. One can identify the weak points by focusing on the dislikes, negative comments, and identifying in which task the students abandon your course.

It is also very useful to have a Teaching Assistant that can follow the course Forum daily to answer student questions as soon as possible. But instructors should also encourage that students themselves try to help their peers in the Forum. In addition, each time a question is made in the Forum because something in the course is not well designed, the instructor should try to think how to change the course material so that that doubt do not show up once more. In this way, the course is refined over time and the number of student doubts due to defects in the course design decrease with time, making the course more effective and giving less work to the instructors and teaching assistants.

Teachers and educators' willingness to prepare and offer rich educational experiences through MOOCs must almost compulsorily use and benefit from the technological infrastructure and promotion channels that the business models of online education companies have. However, the configuration of the business models of online education companies can present relevant ambiguities when it comes to considering that they are structured in formats of digital ecosystems. Such a structure must necessarily contain interests that are not always convergent, a characteristic common to the models of platforms and business ecosystems. The maturation of these ecosystems and the consolidation of the still unstable business models should contribute to reducing the divergences in fields such as those of the commercial objectives on which companies are centered and the educational purpose characteristic of universities and other institutions of education. This is also a manifestation, in the online world, of ``classic'' dilemmas that have always involved the interaction between educational and business objectives.

\bibliographystyle{sbc}
\bibliography{sbc-template}

\begin{thebibliography}{}

\bibitem[Chesbrough 2010]{chesbrough2010business}
Chesbrough, H. (2010).
\newblock Business model innovation: opportunities and barriers.
\newblock {\em Long range planning}, 43(2-3):354--363.

\bibitem[Clift et~al. 2016]{clift2016educational}
Clift, E., Liptak, V., and Rosen, D. (2016).
\newblock Educational ideas and the future of higher education: The quest for a
  new business model.
\newblock In {\em Von der Kutsche zur Cloud--globale Bildung sucht neue Wege},
  pages 7--37. Springer.

\bibitem[Daniel 2012]{Daniel2012}
Daniel, J. (2012).
\newblock {Making Sense of MOOCs: Musings in a Maze of Myth, Paradox and
  Possibility}.
\newblock {\em Journal of Interactive Media in Education}, 3.

\bibitem[Epelboin 2017]{epelboin2017moocs}
Epelboin, Y. (2017).
\newblock Moocs: A viable business model?
\newblock In {\em Open Education: from OERs to MOOCs}, pages 241--259.
  Springer.

\bibitem[Ferdig et~al. 2020]{ferdig2020teaching}
Ferdig, R.~E., Baumgartner, E., Hartshorne, R., Kaplan-Rakowski, R., and Mouza,
  C. (2020).
\newblock Teaching, technology, and teacher education during the covid-19
  pandemic: Stories from the field.
\newblock {\em Waynesville, NC, USA: Association for the Advancement of
  Computing in Education (AACE)}.

\bibitem[Guo 2017]{guo2017mooc}
Guo, P. (2017).
\newblock Mooc and spoc, which one is better?
\newblock {\em Eurasia Journal of Mathematics, Science and Technology
  Education}, 13(8):5961--5967.

\bibitem[Guo et~al. 2014]{guo2014video}
Guo, P.~J., Kim, J., and Rubin, R. (2014).
\newblock {How video production affects student engagement: An empirical study
  of MOOC videos}.
\newblock In {\em Proceedings of the first ACM conference on Learning at scale
  conference}, pages 41--50.

\bibitem[Jansen et~al. 2017]{jansen2017quality}
Jansen, D., Rosewell, J., and Kear, K. (2017).
\newblock Quality frameworks for moocs.
\newblock In {\em Open Education: from OERs to MOOCs}, pages 261--281.
  Springer.

\bibitem[Jung and Lee 2018]{jung2018learning}
Jung, Y. and Lee, J. (2018).
\newblock Learning engagement and persistence in massive open online courses
  (moocs).
\newblock {\em Computers \& Education}.

\bibitem[Kalman 2014]{kalman2014race}
Kalman, Y.~M. (2014).
\newblock A race to the bottom: Moocs and higher education business models.
\newblock {\em Open Learning: The Journal of Open, Distance and e-Learning},
  29(1):5--14.

\bibitem[Osterwalder and Pigneur 2010]{osterwalder2010business}
Osterwalder, A. and Pigneur, Y. (2010).
\newblock {\em Business model generation: a handbook for visionaries, game
  changers, and challengers}.
\newblock John Wiley \& Sons.

\bibitem[Phothilimthana and Sridhara 2017]{Phothilimthana2017}
Phothilimthana, P. M.~. and Sridhara, S. (2017).
\newblock {High-Coverage Hint Generation for Massive Courses}.
\newblock In {\em 22nd Annual Conference on Innovation and Technology in
  Computer Science Education (ITiCSE)}.

\bibitem[Redmond et~al. 2018]{redmond2018online}
Redmond, P., Abawi, L.-A., Brown, A., Henderson, R., and Heffernan, A. (2018).
\newblock An online engagement framework for higher education.
\newblock {\em Online Learning}, 22(1):183--204.

\bibitem[Rothe et~al. 2018]{rothe2018competition}
Rothe, H., Tauscher, K., and Basole, R.~C. (2018).
\newblock {COMPETITION BETWEEN PLATFORM ECOSYSTEMS: A LONGITUDINAL STUDY OF
  MOOC PLATFORMS}.
\newblock In {\em Twenty-Sixth European Conference on Information Systems
  (ECIS2018)}.

\bibitem[Savery 2010]{Savery2010}
Savery, J.~R. (2010).
\newblock {BE VOCAL : Characteristics of Successful Online Instructors}.
\newblock {\em Journal of Interactive Online Learning}, 9(2):141--152.

\bibitem[Sinclair and Kalvala 2016]{sinclair2016student}
Sinclair, J. and Kalvala, S. (2016).
\newblock Student engagement in massive open online courses.
\newblock {\em International Journal of Learning Technology}, 11(3):218--237.

\bibitem[Stephens-Martinez and Fox 2018]{Fox2018}
Stephens-Martinez, K. and Fox, A. (2018).
\newblock {Giving Hints is Complicated: Understanding the challenges of an
  automated hint system based on frequent wrong answers}.
\newblock In {\em 23rd Annual Conference on Innovation and Technology in
  Computer Science Education (ITiCSE)}.

\bibitem[Teece 2010]{teece2010business}
Teece, D.~J. (2010).
\newblock Business models, business strategy and innovation.
\newblock {\em Long range planning}, 43(2-3):172--194.

\bibitem[Wendler et~al. 2017]{wendler2017business}
Wendler, W.~S., Stumpf-Wollersheim, J., and Welpe, I.~M. (2017).
\newblock Business models in the education technology industry: What makes them
  successful?
\newblock In {\em International Conference on Information Systems (ICIS)}.

\bibitem[Yang et~al. 2018]{yang2018online}
Yang, D., Lavonen, J.~M., and Niemi, H. (2018).
\newblock Online learning engagement: Factors and results-evidence from
  literature.
\newblock {\em Themes in eLearning}, 11(1):1--22.

\bibitem[Yang et~al. 2017]{yang2017moocs}
Yang, S.~J., Huang, J.~C., and Huang, A.~Y. (2017).
\newblock Moocs in taiwan: The movement and experiences.
\newblock In {\em Open Education: from OERs to MOOCs}, pages 101--116.
  Springer.

\end{thebibliography}
%\nocite{Baker2011}
%\nocite{Seffrin2013}
%\nocite{Brasil2008}
%\nocite{Kautzman2015}
%\nocite{Sweller1991}
%\nocite{Clark2006}
%\nocite{Mason2012}

\end{document}